\documentclass[conference,twoside,letterpaper,10pt]{IEEEtran}
\IEEEoverridecommandlockouts

\ifCLASSINFOpdf
\else
\fi

\usepackage[cmex10]{amsmath}
\usepackage{cite}
\usepackage{mathtools}
\usepackage{url}
\usepackage{flushend}
\usepackage{enumerate}
\begin{document}
%
\title{Feasibility of In-band Full-Duplex Radio Transceivers with Imperfect RF Components:\\ Analysis and Enhanced Cancellation Algorithms\vspace{-3mm}}

\author{\IEEEauthorblockN{Dani~Korpi,
Lauri~Anttila,
and~Mikko~Valkama\vspace{-2mm}}
\\
\IEEEauthorblockA{Department of Electronics and Communications Engineering, Tampere University of Technology, Finland\\ e-mail: dani.korpi@tut.fi, lauri.anttila@tut.fi, mikko.e.valkama@tut.fi}
\thanks{The research work leading to these results was funded by the Academy of Finland (under the project \#259915 "In-band Full-Duplex MIMO Transmission: A Breakthrough to High-Speed Low-Latency Mobile Networks"), the Finnish Funding Agency for Technology and Innovation (Tekes, under the project "Full-Duplex Cognitive Radio"), the Linz Center of Mechatronics (LCM) in the framework of the Austrian COMET-K2 programme, and Emil Aaltonen Foundation.}}%

\maketitle

\begin{abstract}
In this paper we provide an overview regarding the feasibility of in-band full-duplex transceivers under imperfect RF components. We utilize results and findings from the recent research on full-duplex communications, while introducing also transmitter-induced thermal noise into the analysis. This means that the model of the RF impairments used in this paper is the most comprehensive thus far. By assuming realistic parameter values for the different transceiver components, it is shown that IQ imaging and transmitter-induced nonlinearities are the most significant sources of distortion in in-band full-duplex transceivers, in addition to linear self-interference. Motivated by this, we propose a novel augmented nonlinear digital self-interference canceller that is able to model and hence suppress all the essential transmitter imperfections jointly. This is also verified and demonstrated by extensive waveform simulations.
\end{abstract}

\section{Introduction}

Full-duplex radio communications with simultaneous transmission and reception at the same radio frequency (RF) carrier has recently gained considerable interest among researchers. It has the potential to significantly improve the efficiency and flexibility of RF spectrum usage, which makes it an appealing concept when trying to increase the data rates of the current systems, while retaining the amount of utilized resources. There have already been several promising implementations of such full-duplex radio transceivers \cite{Choi10,Jain11,Duarte10,Bharadia13}. In addition to practical demonstrations, there has also been a large number of theoretical studies investigating the boundaries of in-band full-duplex communications under various impairments and circumstances \cite{Korpi133,Korpi13,Riihonen122}.


The most crucial issue in wireless single channel full-duplex communications is the own transmit signal that is coupled back to the receiver and acts as a strong source of interference. This so-called \textit{self-interference} (SI) can be as much as 60--100 dB more powerful than the weak received signal of interest, and thus it must be attenuated significantly to allow the detection of the actual received signal. Typically the attenuation of the SI signal is done in two stages: first at the input of the receiver (RX) chain, and then after the analog-to-digital conversion \cite{Choi10,Jain11,Duarte10,Bharadia13}. These SI cancellation stages are usually referred to as \textit{RF cancellation} and \textit{digital cancellation}, respectively. RF cancellation is required in order to prevent the complete saturation of the receiver components and the analog-to-digital converter (ADC). Digital cancellation is then used to attenuate the rest of the SI signal below the noise floor. Both of these cancellation methods rely on processing the known transmit signal to produce the cancellation signal.


As the research on single channel full-duplex communications is still at a rather early stage, it is important to determine its overall feasibility under practical considerations. In the context of wireless communications, this means in particular that the effects of the different non-idealities must be rigorously analyzed. Actually, the RF impairments occurring within in-band full-duplex transceivers are the most significant challenge for its practical implementation, as the basic concept of self-interference cancellation is very simple. Namely, in the ideal case it is sufficient to estimate the linear channel experienced by the SI signal and then generate a corresponding cancellation signal to be subtracted from the received signal. However, the analog impairments may prevent the usage of such a simple procedure, and thus form a substantial challenge in achieving a sufficient amount of SI attenuation.

In this paper, we provide a thorough overview on the feasibility of wireless in-band full-duplex communications. We will go through the most prominent types of analog impairments, and also some methods for suppressing them, including a novel joint cancellation algorithm capable of attenuating the most prominent distortion components. In addition, we will also consider the combined effect of the discussed impairments to provide some insight into the achievable performance of an in-band full-duplex transceiver. In this analysis, the transmitter (TX)-induced thermal noise will also be taken into account, which is something that has not been done very systemically before to the best of our knowledge. Hence, the feasibility analysis presented in this paper is the most comprehensive one presented thus far.

The rest of this paper is organized as follows. In Section~\ref{sec:rf_imp}, we will briefly go through some of the most prominent RF impairments occurring in in-band full-duplex transceivers. Then, in Section~\ref{sec:canc_alg}, some previously proposed methods for compensating these impairments are discussed, alongside with the proposed novel joint digital cancellation scheme. In Section~\ref{sec:simul} we compare the performance of these algorithms by means of very comprehensive waveform simulations. Finally, the conclusions are drawn in Section~\ref{sec:conc}.

\emph{Nomenclature:} Throughout the paper, the use of linear power units is indicated by lowercase letters. Correspondingly, when referring to logarithmic power units, uppercase letters are used. The only exception to this is the noise factor, which is denoted by capital $F$ according to common convention in the literature of the field. Watts are used as the absolute power unit, and dBm as the logarithmic power unit.

\section{RF Impairments in Full-duplex Radio Transceivers}
\label{sec:rf_imp}

In practical transceivers, none of the components are completely ideal, which means that they will distort the signal in numerous ways. In half-duplex communications, a relatively high level of distortion can be tolerated, as the transmitted and received signals are well separated. However, this is not the case for in-band full-duplex transceivers, where the transmitted and received signals overlap freely. Due to the large difference in the powers of the transmitted signal and the received signal of interest, especially when operating close to the sensitivity level of the receiver, even relatively mild distortion of the overall signal may lead to a drastic decrease in the final signal-to-interference-plus-noise ratio (SINR).

\subsection{Nonlinear Distortion}

Perhaps the most widely studied source of distortion is the nonlinearity of the amplifiers. Mostly, only the nonlinearity of the transmitter power amplifier (PA) has been considered, as it distorts the actual transmitted signal and thus leads to residual self-interference (SI) if employing only linear cancellation techniques \cite{Korpi13,Anttila13,Ahmed13,Bharadia13}. However, it has also been shown that the RX chain can contribute significantly to the nonlinear distortion of the SI signal \cite{Korpi13,Korpi132}. The reason for this is that typically the receiver components have been dimensioned for the relatively weak signal of interest, and thus the powerful SI signal will force the amplifiers into their nonlinear operating region. This applies particularly to the baseband amplification stages.

It is obviously possible to combat the effect of nonlinear distortion in full-duplex transceivers by opting for more linear transceiver components. However, while this would decrease the level of distortion, it would also be significantly more expensive. For this reason, different signal processing techniques have been proposed for attenuating nonlinearly distorted SI signals \cite{Anttila13,Ahmed13,Bharadia13,Korpi132}. The basic idea behind these techniques is to model the nonlinearity of the amplifiers with a polynomial model, possibly with memory, and then estimate the coefficients of these polynomials. This is relatively straight-forward when assuming that only the PA is producing significant nonlinear distortion, but the procedure is slightly more complicated when considering also the nonlinearity of the RX chain \cite{Anttila13,Korpi132}. Overall, these nonlinear SI cancellation methods allow for compensating the nonlinearly distorted SI signal in the digital domain, based only on the known transmit samples. All the necessary information can be estimated during a calibration period, as long as the model for the overall nonlinearities is known and fixed.

Some insight regarding the power level of the nonlinear distortion can be obtained with system calculations. A well-known model for the power of an $n$th order nonlinearity is defined as
\begin{align}
	P_{nth} = P_{out} - (n-1)(\mathit{IIPn}-P_{in}) \text{,} \label{eq:nl_basic}
\end{align}
where $P_{out}$ is the power of the fundamental signal at the output of the component, $\mathit{IIPn}$ is the $n$th order input-referred intercept point, and $P_{in}$ is the power of the fundamental signal at the input of the component \cite{Gu06}. The above equation assumes the signal powers to be in dBm units. Using \eqref{eq:nl_basic}, it is possible to approximate the power of the nonlinear distortion in comparison to the power levels of the other signal components. For a detailed analysis, refer to \cite{Korpi13}.

\subsection{IQ Imaging}
\vspace{-0.5mm}

Another significant issue in most wireless radio transceivers is IQ imaging \cite{Anttila11}, which is caused by the mismatches between the I- and Q-branches of the TX and RX chains. This is a widespread issue in wireless communications since nowadays most transceiver structures utilize IQ processing. It can be shown that the imbalance between the I- and Q-branches results in the complex conjugate of the ideal signal being summed on top of it with certain attenuation. Thus, for an input signal $x(t)$, the output of an imperfect IQ mixer is of the following form:
\begin{align}
	x_{IQ}(t) = g_{1}(t) \star x(t) + g_{2}(t) \star x^{\ast}(t) \text{,} \label{eq:iq_imb}
\end{align}
where $g_{1}(t)$ and $g_{2}(t)$ are the responses for the direct signal component and the image component, respectively \cite{Anttila11}. Here $(\cdot)^{\ast}$ indicates the complex conjugate and $\star$ denotes the convolution operation. The quality of the IQ mixer can be quantified with image rejection ratio (IRR), which is defined as
\begin{align}
	\mathit{IRR}(f) = 10\log_{10}\left(\frac{|G_{1}(f)|^2}{|G_{2}(f)|^2}\right) \text{,} \label{eq:irr}
\end{align}
where $G_{1}(f)$ and $G_{2}(f)$ are the frequency-domain representations of $g_{1}(t)$ and $g_{2}(t)$, respectively \cite{Anttila11}. Based on \eqref{eq:irr}, the power level of the IQ image component can be approximated and used in a system calculations based feasibility analysis.

The effect of IQ imbalance on in-band full-duplex transceivers is studied in detail in \cite{Korpi133}. It is shown that, assuming a practical IQ image rejection ratio for a full-duplex transceiver, it is necessary to attenuate also the IQ image of the SI signal. Otherwise the loss of SINR might be in the order of tens of decibels, and there is no performance gain due to simultaneous transmission and reception on the same frequency-band. The authors then propose a novel digital cancellation scheme referred to as widely-linear digital cancellation, which allows the modeling of the effects of IQ imbalance and, consequently, also the compensation of the SI mirror image. A significant increase in the achievable SINR is reported when using the proposed scheme. Thus, the results and analysis shown in \cite{Korpi133}, alongside with the observations made in \cite{Hua12}, indicate that IQ imbalance is a serious concern in the context of in-band full-duplex transceivers, and it must be included in the discussion regarding the feasibility of single-channel full-duplex communications.


\vspace{-1.5mm}
\subsection{Quantization Noise}

Due to the highly powerful SI signal, the dynamic range of the ADC is also a key concern in in-band full-duplex transceivers \cite{Korpi13,Riihonen122}. Namely, if the SI signal is not attenuated sufficiently in the analog domain, it will reserve some or even most of the dynamic range of the ADC, which means that the resolution of the signal of interest will be very low. This results in a decreased SINR at the detector due to the quantization noise floor.

The absolute power level of the quantization noise floor in logarithmic power units can be calculated as
\begin{align}
	P_q = P_{AD} - \mathit{SNR}_{ADC} \text{,} \label{eq:p_q}
\end{align}
where $P_{AD}$ is the total power of the signal at the input of the analog-to-digital converter (ADC) and $\mathit{SNR}_{ADC}$ is the signal-to-(quantization)-noise ratio of the ADC. The value of the ADC SNR can be calculated as $\mathit{SNR}_{ADC} = 6.02b+4.76-\mathit{PAPR}$, where $b$ is the number of bits and $\mathit{PAPR}$ is the peak-to-average-power ratio of the total signal \cite{Gu06}. Assuming an ideal automatic gain control algorithm in the receiver, higher SI power means that the power of the desired signal is lower, which results in the quantization noise floor being closer to the power level of the desired signal. This is one way of expressing the decrease in the bit resolution of the desired signal. Using \eqref{eq:p_q}, this decrease can then be approximated with realistic numerical values.

It is possible to combat against this issue in several ways. Perhaps the easiest and most straightforward method is to increase the number of bits at the ADC. It is shown, for example, in \cite{Korpi13} that each additional bit increases the dynamic range of the ADC by approximately 6 dB. Of course, there are practical limitations on the maximum number of bits, but it is recommendable to have as much bits as possible in an in-band full-duplex transceiver to ensure a sufficient resolution for the signal of interest in the presence of powerful SI.

A more sophisticated method for combatting the quantization noise is to increase the amount of SI attenuation before the ADC. This can be done either by improving the performance of the RF cancellation, or by implementing an additional analog baseband SI cancellation stage before the analog-to-digital conversion. In \cite{Kaufman13}, the authors used the latter technique to decrease the burden of the ADC. When assessing the performance of the analog baseband cancellation scheme, it was observed that the additional cancellation stage resulted in a higher SINR than could be achieved without it.

\vspace{-0.5mm}
\subsection{Transmitter-induced Thermal Noise}

One aspect of wireless in-band full-duplex transceivers, which has been largely neglected in earlier literature, is the thermal noise occurring within the transmitter and RF cancellation chains. In typical half-duplex applications, this TX-induced thermal noise is not a concern, as it is very low in comparison to the other components of the transmit signal. However, in full-duplex transceivers, the power of the TX-induced thermal noise might not always be negligibly low, as the power of the transmit signal is very high in comparison to the power of the weak received signal of interest. Thus, it might increase the overall noise floor if the separation between the TX and RX chains is not sufficiently high. For this reason, we will consider also the thermal noise produced by the transmitter in the subsequent analysis. Although there have been some studies where also the baseband cancellation signal has been chosen such that it includes all the noise components produced in the transmit path \cite{Hua13}, to the best of our knowledge this is the first explicit analysis regarding TX-induced thermal noise in full-duplex context.

\begin{figure*}[!t]
\centering
\includegraphics[width=0.75\textwidth]{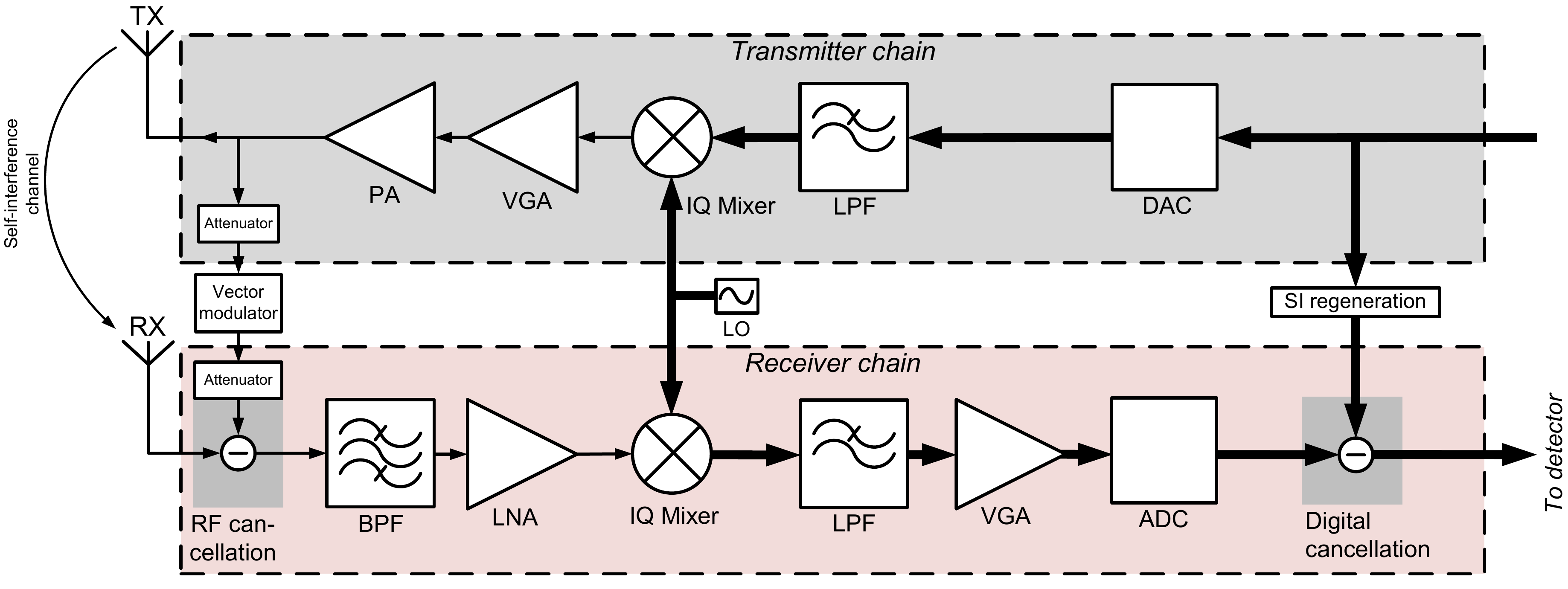}
\caption{A block diagram of the considered direct-conversion full-duplex transceiver.}
\label{fig:block_diagram}
\vspace{-5mm}
\end{figure*}

As a starting point for the analysis, let us consider a direct-conversion full-duplex transceiver, whose structure is as shown in Fig.~\ref{fig:block_diagram}. At the output of the digital-to-analog converter (DAC), the power of the thermal noise is obviously at the level of the thermal noise floor. Assuming an ideal low-pass filter (LPF), the power of the thermal noise is increased only by the IQ mixer, transmitter variable gain amplifier (VGA), and the PA. The TX-induced thermal noise is then coupled to the RX chain through the RF cancellation path. There, the level of the amplified thermal noise is decreased because the reference signal must be heavily attenuated to match its power level to that of the actual SI signal. Additional noise is also produced in the RF cancellation path, as an active vector modulator (VM) is required to tune the phase and amplitude of the cancellation signal. However, by having attenuators both at the input and output of the VM, the possible additional noise introduced by it can be somewhat attenuated. In practical implementations, the attenuation can be done, e.g., with directional couplers and combiners.

When the cancellation signal is then added to the received signal, the power of the TX-induced thermal noise is further decreased because part of the noise signal is included in both the SI signal and cancellation signal, but obviously with opposite phases. In essence, this means that the thermal noise signal produced in the actual transmitter is attenuated by the same amount as the SI signal is attenuated between the output of the PA and the input of the low-noise amplifier (LNA). Thus, assuming that the additional noise produced by the VM is attenuated sufficiently, in some cases the power of the TX-induced thermal noise will be negligibly low in comparison to the other signal components. However, under some conditions, it might decrease the achievable SINR of a full-duplex transceiver, which is why we will explicitly analyze it in this paper.

To quantify exactly the effect of the TX-induced thermal noise, let us derive an expression for the overall thermal noise power at the input of the receiver detector, including both the TX- and RX-induced contributions to the thermal noise level. Assuming a similar structure as shown in Fig.~\ref{fig:block_diagram} and taking into account the assumptions discussed above, the approximative power of the thermal noise at the receiver detector input can be shown to be
\begin{align}
	{p}_{n} &= \left| {{k}_{BB}} \right|^{2} \left| {{k}_{LNA}} \right|^{2}\left| {{g}_{1,RX}} \right|^{2} \left[ F +\left|a_2 \right|^2 (\left|a_{vm} \right|^2 F_{vm}-1)\right. \nonumber\\
&\left. {}-{\left| {{a}_{ant}} \right|}^{2}+\left| {{a}_{ant}} \right|^{2} \left| {{a}_{RF}} \right|^{2} \left| {{k}_{PA}} \right|^{2} \left( F_{PA}-1 \right.\right. \nonumber\\
&\left.\left. {} +F_{tx}\left| {{g}_{1,TX}} \right|^{2}\left| {{k}_{VGA}} \right|^{2} \right) \right]{p}_{th}\text{,} \label{eq:p_noise}
\end{align}
where $\left|{k}_{BB}\right|^{2}$ is the gain of the receiver VGA, $\left|{k}_{LNA}\right|^{2}$ is the gain of the LNA, $\left|{g}_{1,RX}\right|^{2}$ is the gain of the direct path in the receiver IQ mixer, $F$ is the total noise factor of the receiver chain, $\left| {{a}_{2}} \right|^{2}$ is the amount of attenuation done after the VM, $\left| {{a}_{vm}} \right|^{2}$ is the (negative) gain of the VM, $F_{vm}$ is the noise factor of the VM, $\left| {{a}_{ant}} \right|^{2}$ is the amount of antenna separation, $\left| {{a}_{RF}} \right|^{2}$ is the amount of RF cancellation, $\left| {{k}_{PA}} \right|^{2}$ is the gain of the PA, ${F}_{PA}$ is the noise factor of the PA, ${F}_{tx}$ is the combined noise factor of the transmitter VGA and IQ mixer, $\left|{g}_{1,TX}\right|^{2}$ is the gain of the direct path in the transmitter IQ mixer, $\left| {{k}_{VGA}} \right|^{2}$ is the gain of the transmitter VGA, and $p_{th}$ is the power of the thermal noise floor.


The above equation assumes that all the responses of the transceiver components are frequency independent. This is a justified approximation in this context as we are only interested in the average power levels of the different signal components. We have also neglected the IQ imaging of the thermal noise signal, as it has no significant effect on its final power level. The actual derivation of \eqref{eq:p_noise} has been done based on a similar signal model as presented in \cite{Korpi133}, and further details regarding the derivation process can be read from there. In this paper, we must omit the derivation of \eqref{eq:p_noise} for brevity. Also note that this equation is evaluated using linear power units, which is also indicated by the lowercase letters.


If the objective is to analyze the effect of TX-induced thermal noise separately, \eqref{eq:p_noise} can be rewritten as
\begin{align}
	{p}_{n} &= p_{n,RX}+p_{n,TX} = \left| {{k}_{BB}} \right|^{2} \left| {{k}_{LNA}} \right|^{2}\left| {{g}_{1,RX}} \right|^{2} F {p}_{th} \nonumber\\
	&+ \left| {{k}_{BB}} \right|^{2} \left| {{k}_{LNA}} \right|^{2}\left| {{g}_{1,RX}} \right|^{2}\left[\left|a_2 \right|^2 (\left|a_{vm} \right|^2 F_{vm}-1)\right. \nonumber\\
&\left. {} - {\left| {a_{ant}} \right|}^2 + {\left| {{a}_{ant}} \right|}^{2} \left| {{a}_{RF}} \right|^{2} \left| {{k}_{PA}} \right|^{2} \left( {F}_{PA}-1\right.\right. \nonumber\\
&\left.\left. {}+{F}_{tx}\left| {{g}_{1,TX}} \right|^{2}\left| {{k}_{VGA}} \right|^{2} \right) \right]{p}_{th}\text{,} \label{eq:p_noise_sep}
\end{align}
where the first term represents the RX-induced thermal noise ($p_{n,RX}$) and the latter term represents the TX-induced thermal noise ($p_{n,TX}$), respectively. Using either \eqref{eq:p_noise} or \eqref{eq:p_noise_sep} it is then possible to calculate the power level of the overall thermal noise at the receiver detector input of an in-band full-duplex transceiver, taking into account also the TX-induced thermal noise.

\vspace{-2mm}
\subsection{Overall Feasibility}

To determine the overall feasibility of wireless single channel full-duplex communications, let us again consider a typical full-duplex transceiver with a structure shown in Fig.~\ref{fig:block_diagram}. The considered full-duplex transceiver follows a typical direct-conversion architecture, and it has two self-interference cancellation stages, namely RF cancellation and digital cancellation, as already discussed earlier \cite{Choi10,Jain11,Duarte10,Korpi13}. In this section, we will determine the relative power levels of the different signal components with realistic transceiver component parameters. The power levels are calculated with simplified system calculations, based on \eqref{eq:nl_basic}, \eqref{eq:irr}, \eqref{eq:p_q}, and \eqref{eq:p_noise_sep}. However, for brevity, we dont derive the actual equations for the accumulated power levels in this paper, but use the equations derived in \cite{Korpi133} and \cite{Korpi13}, where a similar in-band full-duplex transceiver is analyzed. The power of the thermal noise, on the other hand, is calculated with \eqref{eq:p_noise_sep}, as it has not been derived in any of the earlier studies.

Next, let us define realistic parameters values for the different components. All the parameters are listed in Tables~\ref{table:system_parameters} and~\ref{table:rx_parameters}, and they have been chosen based on earlier literature and LTE specifications to represent a realistic in-band full-duplex transceiver \cite{Parssinen99,Behzad07,Gu06,Jain11,Choi10,LTE_specs}. Note that in the transmitter it is assumed that the IQ mixer and the VGA are linear, meaning that only the PA produces nonlinear distortion to the transmit signal. In addition, to simplify the notations, the noise figures of the IQ mixer and VGA are combined, which is indicated in Table~\ref{table:rx_parameters} by the IQ mixer having a noise figure of 10 dB and the VGA having no noise figure at all. The parameters for the VM have been chosen according to \cite{VM_datasheet} with the exception that in this analysis we assume the VM to be linear. It is further assumed that the reference signal is attenuated by 15 dB before and after the VM, which means that the $-10$ dB average gain of the VM will be sufficient match the power of the reference signal to that of the actual SI signal.

\begin{table}[!t]
\renewcommand{\arraystretch}{1.3}
\caption{Baseline system level parameters of the full-duplex transceiver.}
\label{table:system_parameters}
\centering
\begin{tabular}{|c||c|}
\hline
\textbf{Parameter} & \textbf{Value}\\
\hline
SNR requirement & 10 dB \\
\hline
Bandwidth & 12.5 MHz \\
\hline
Receiver noise figure & 4.1 dB\\
\hline
Sensitivity & $-$88.9 dBm\\
\hline
Received signal power & $-$83.9 dBm\\
\hline
Antenna separation & 40 dB\\
\hline
RF cancellation & 30 dB\\
\hline
Attenuation before the VM & 15 dB\\
\hline
Attenuation after the VM & 15 dB\\
\hline
Image rejection ratio (TX and RX) & 30 dB\\
\hline
ADC bits & 12\\
\hline
ADC P-P voltage range & 4.5 V\\
\hline
PAPR & 10 dB\\
\hline
\end{tabular}
\vspace{-2mm}
\end{table}


\begin{table}[!t]
\renewcommand{\arraystretch}{1.3}
\caption{Parameters for the individual components of the full-duplex transceiver.}
\label{table:rx_parameters}
\centering
\begin{tabular}{|c||c||c||c||c|}
\hline
\textbf{Component} & \textbf{Gain [dB]} & \textbf{IIP2 [dBm]} & \textbf{IIP3 [dBm]} & \textbf{NF [dB]}\\
\hline
IQ mixer (TX) & 6 & - & - & 10\\
\hline
VGA (TX) & 0--30 & - & - & -\\
\hline
PA (TX) & 27 & - & 13 & 5\\
\hline
VM & -10 & - & - & 20\\
\hline
LNA (RX) & 25 & 43 & $-$9 & 4.1\\
\hline
IQ mixer (RX) & 6 & 42 & 15 & 4\\
\hline
VGA (RX) & 0--69 & 43 & 14 & 4\\
\hline
\end{tabular}
\vspace{-5mm}
\end{table}

\begin{figure}[!t]
\centering
\includegraphics[width=\columnwidth]{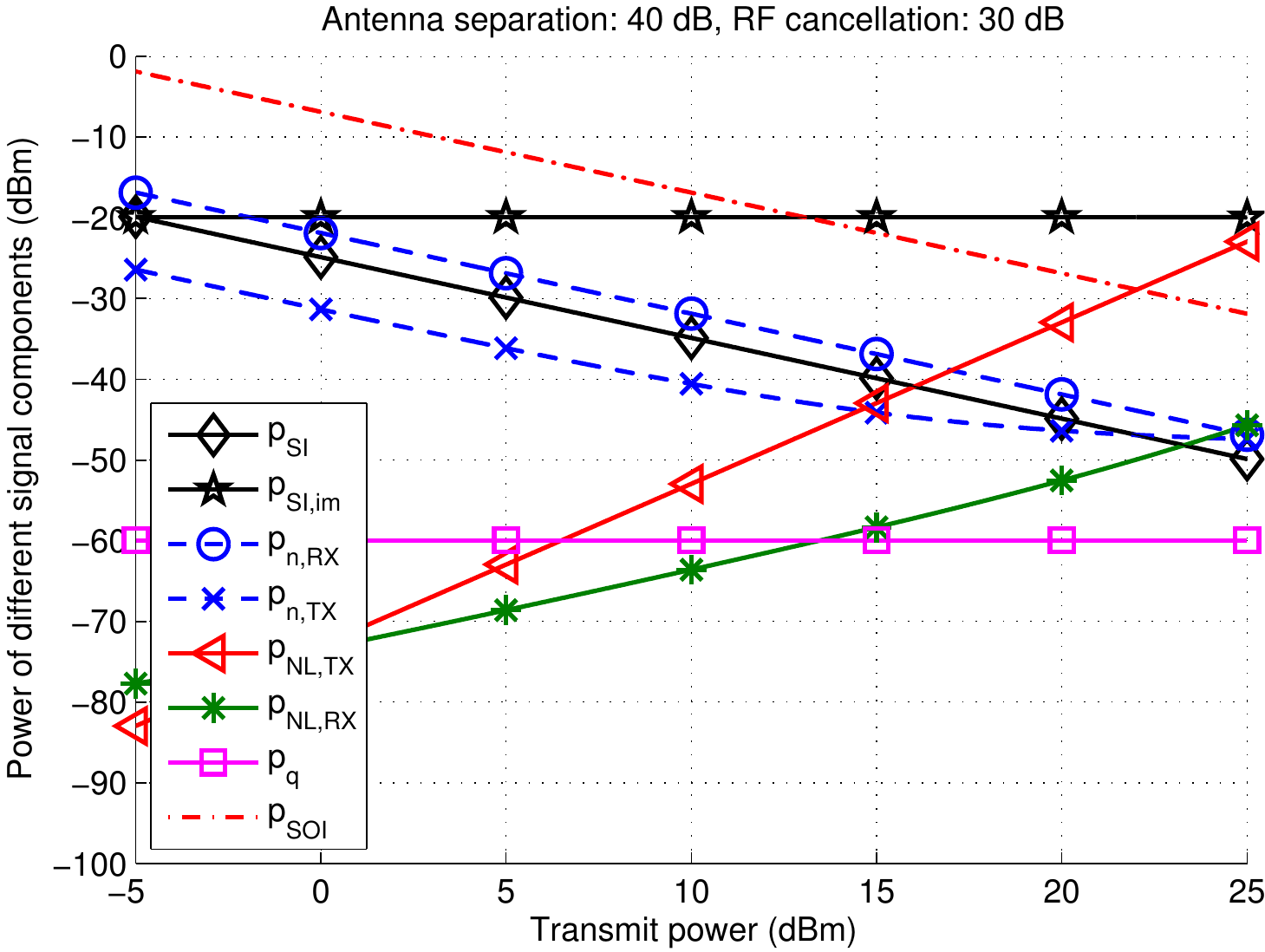}
\caption{Power levels of the different signal components at the input of the receiver detector.}
\label{fig:syscalc}
\vspace{-5mm}
\end{figure}

Having specified the parameters, the power levels of the different signal components can then be approximated using \eqref{eq:p_noise_sep}, and the equations derived in \cite{Korpi133} and \cite{Korpi13}. The resulting signal component powers are shown in Fig.~\ref{fig:syscalc}, and they correspond to a scenario where only linear digital cancellation is performed. In the legend, $p_{SI}$ refers to the linear self-interference signal, $p_{SI,im}$ refers to its mirror image, i.e., the conjugate SI, $p_{n,RX}$ and $p_{n,TX}$ refer to the RX- and TX-induced thermal noise, $p_{NL,TX}$ refers to the PA-induced nonlinear distortion, $p_{NL,RX}$ refers to the nonlinear distortion produced by the receiver components, $p_{q}$ refers to the quantization noise, and $p_{SOI}$ refers to the signal of interest. For these calculations, the amount of linear digital cancellation has been assumed to be such that the linear SI signal ($p_{SI}$) is attenuated slightly below the thermal noise floor. Furthermore, for improved readability, some negligibly weak signal components have been omitted from Fig.~\ref{fig:syscalc}, as they were observed to have no contribution to the overall noise power. These weak signal components include, for instance, the mirror images of the nonlinear distortion and thermal noise. Note that the absolute power levels of most of the signal components are decreasing due to the automatic gain control at the RX chain, which matches the power of the ADC input signal to the available ADC voltage range. With higher transmit powers, less gain is required due to the higher self-interference power.

When investigating Fig.~\ref{fig:syscalc} more closely, it can be observed that conjugate SI ($p_{SI,im}$) is clearly the most powerful signal component after linear digital cancellation. This is a similar observation as done already in \cite{Korpi133}. Thus, it is obvious that the conjugate SI must be attenuated by some means to make in-band full-duplex communications in any way feasible. In addition to IQ imaging, another serious concern for full-duplex transceivers is the nonlinear distortion. From Fig.~\ref{fig:syscalc} it can be seen that, with higher transmit powers, especially the PA-induced nonlinearities will decrease the SINR. Similarly, also the nonlinear distortion produced by the receiver components can be troublesome with very high transmit powers.

Fig.~\ref{fig:syscalc} indicates also that the TX-induced thermal noise is not a serious concern with these parameters, as its level remains relatively low with most of the transmit powers. However, as can be observed from \eqref{eq:p_noise_sep}, the power of the TX-induced thermal noise is dependent on the amount of antenna separation and RF cancellation. Thus, with less efficient analog SI cancellation, it can pose a serious issue for an in-band full-duplex transceiver. Nevertheless, with the considered parameters, TX-induced thermal noise is not among the most prominent sources of distortion. Finally, even though not considered here, we wish to emphasize that also oscillator phase noise may pose a performance limit for in-band full-duplex transceivers. In this paper, however, it is assumed that the TX and RX signals have a common local oscillator signal, which means that the level of the phase noise remains on a tolerable level \cite{Syrjala13}. For more details, refer to \cite{Syrjala13,Sahai12}.


\vspace{-1.5mm}
\section{Enhanced Cancellation Algorithms for Inband Full-Duplex Radio Transceivers}
\label{sec:canc_alg}

As already discussed in Section~\ref{sec:rf_imp}, several compensation methods for different RF impairments occurring in full-duplex transceivers have been developed. Based on the observation from Fig.~\ref{fig:syscalc} that conjugate SI and PA-induced nonlinearies are the most prominent sources of distortion when assuming typical parameters, we will briefly present some enhanced cancellation algorithms for attenuating these nonidealities.

\vspace{-2mm}
\subsection{Widely-linear Digital Cancellation}
\label{sec:wl_canc}

As already shown in \eqref{eq:iq_imb}, IQ imbalance will result in the complex conjugate of the original signal being summed on top of it. In \cite{Korpi133} it is shown that this results in the SI signal being of the following form in the digital domain:
\begin{align}
	y_{ADC}(n) = h_1(n) \star x(n) + h_2(n) \star x^{\ast}(n) + z(n) \text{,} \label{eq:iq_signal}
\end{align}
where $x(n)$ is the original transmitted signal, $x^{\ast}(n)$ is its complex conjugate, $h_1(n)$ and $h_2(n)$ are the respective total responses of the linear SI and conjugate SI, and $z(n)$ denotes the other noise components. This type of a signal model, where both direct and complex-conjugated signals are filtered and finally summed together, is typically called widely-linear in the literature \cite{Picinbono95}.

As the transmitted signal is obviously known within the transceiver, it is sufficient to obtain estimates for the two channel responses, $h_1(n)$ and $h_2(n)$, to produce a widely-linear cancellation signal \cite{Korpi133}. By rewriting \eqref{eq:iq_signal} with vector-matrix notation, and using least-squares, an augmented estimate for these channel responses can be calculated as
\begin{align}
	\mathbf{\hat{h}}_{aug} = (\mathbf{X}_{aug}^{H} \mathbf{X}_{aug})^{-1} \mathbf{X}_{aug}^{H} \mathbf{y}_{ADC} \text{,} \label{eq:h_est}
\end{align}
where $\mathbf{X}_{aug}$ is the augmented data convolution matrix, which includes also the complex conjugated signal elements, $()^{H}$ denotes the Hermitian transpose, and $\mathbf{y}_{ADC}$ contains the samples of \eqref{eq:iq_signal}. For more detailed information regarding the structure of the vectors and matrices, refer to \cite{Korpi133}.

The actual cancellation signal can then be produced by first noting that $\mathbf{\hat{h}}_{aug} = \begin{bmatrix} \mathbf{\hat{h}}_1^T & \mathbf{\hat{h}}_2^T \end{bmatrix}^T$, meaning that
\begin{align*}
	\mathbf{\hat{h}}_{1} &= \begin{bmatrix}\mathbf{\hat{h}}_{aug}(0) & \mathbf{\hat{h}}_{aug}(1) & \cdots & \mathbf{\hat{h}}_{aug}(M-1)\end{bmatrix}^T\\
	\mathbf{\hat{h}}_{2} &= \begin{bmatrix}\mathbf{\hat{h}}_{aug}(M) & \mathbf{\hat{h}}_{aug}(M+1) & \cdots & \mathbf{\hat{h}}_{aug}(2M-1)\end{bmatrix}^T
\end{align*}
based on which the cancellation signal can then be written as
\begin{align*}
	y_{canc}(n) = \hat{h}_1(n) \star x(n) + \hat{h}_2(n) \star x^{\ast}(n) \text{.}
\end{align*}
Here $\hat{h}_1(n)$ and $\hat{h}_2(n)$ denote the impulse responses corresponding to the vectors $\mathbf{h}_1$ and $\mathbf{h}_2$. Finally, the widely-linear digital cancellation procedure can be carried out as
\begin{align*}
	y_{WLDC}(n) &= y_{ADC}(n) - y_{canc}(n)\text{.}
\end{align*}
In \cite{Korpi133} it was observed that this type of a cancellation procedure significantly improves the final SINR with practical IRR values for the in-band full-duplex transceiver. We will also confirm this later in this paper using even more comprehensive modeling in the simulations.

\subsection{Nonlinear Digital Cancellation}
\label{sec:nl_canc}

Compensating for the PA-induced nonlinear distortion is not quite as straight-forward as compensating for IQ imaging, but similar principles can be utilized in the cancellation processing. An enhanced digital cancellation procedure, capable of attenuating also nonlinearly distorted SI signals, is proposed in \cite{Anttila13}. There, a parallel Hammerstein (PH) model is assumed for the transmitter PA. Denoting again the original digital baseband transmit signal by $x(n)$, the output signal of the PA can be written with the PH model as
\begin{align}
x_{PA}(n) = \sum_{\substack{
   p = 1 \\
   p \text{ } \mathit{odd}
  }}^{P}
 \sum_{{k} = 0}^{{M_{PA}}-1} f_{{p}}(k) \psi_{p}(x({n}-{k})) \text{,} \label{eq:ph_model}
\end{align}
where the basis functions are defined as $\psi_{p}(x(n)) = |x(n)|^{{p}-1} x(n)$, $f_{p}(n)$ are FIR filter impulse responses of the PH branches, $M_{PA}$ denotes the memory length, and $P$ denotes the nonlinearity order of the PH model \cite{Isaksson06}. The overall model for the SI signal is also affected by the coupling channel, RF cancellation, and receiver processing. However, as was already noted in \cite{Anttila13}, these transformations do not affect the overall signal model, which still follows the PH structure but just with modified filters $f_{{p,\mathit{eff}}}(k)$, whose lengths might be slightly increased in comparison to $f_{{p}}(k)$, depending on the delay spread of the SI channel. The final cancellation signal is then constructed by estimating the impulse responses of the different effective PH branches, namely $f_{{p,\mathit{eff}}}(k)$ for all $p = \{1,3,\ldots,P \}$.

The actual estimation can be done, for example, with least-squares. Following the derivations in \cite{Anttila13}, the least-squares solution for all $f_{p,\mathit{eff}}(n)$, or $\mathbf{f}_\mathit{eff}$ with augmented vector notation, can be obtained as
\begin{align}
	\mathbf{\hat{f}}_\mathit{eff} = (\mathbf{\Psi}^H \mathbf{\Psi})^{-1} \mathbf{\Psi}^H \mathbf{y}_{ADC}\text{,} \label{eq:nl_canc}
\end{align}
where $\mathbf{\Psi}$ is an augmented convolution matrix constructed with the basis functions $\psi_{p}(x(n))$, and $\mathbf{y}_{ADC}$ is the received signal in the digital domain. For more detailed information regarding \eqref{eq:nl_canc} and its derivation, see \cite{Anttila13}. The actual cancellation signal can then be produced by using the estimated $\mathbf{\hat{f}}_\mathit{eff}$ in PH processing to create a nonlinear SI estimate, and substracting the resulting signal from $\mathbf{y}_{ADC}$. In the next section we will evaluate also the performance of this type of a nonlinear cancellation procedure with a comprehensive simulation model. Note that similar nonlinear cancellation solutions have been reported also in \cite{Bharadia13,Ahmed13}.


\vspace{-1mm}
\subsection{Proposed Joint Augmented Cancellation of Nonlinear Distortion and Conjugate Self-interference}
\label{sec:joint_canc}

In \cite{Korpi133} it was observed that, even though the conjugate SI is typically the dominant distortion component, the PA-induced nonlinearities are also harmful with higher transmit powers, as Fig.~\ref{fig:syscalc} suggests. In fact, they were observed to be the limiting factor for the SINR after widely-linear digital cancellation. For this reason, we will now propose a novel joint cancellation algorithm for attenuating both nonlinear and conjugate SI. The joint cancellation can be done by augmenting the basis function matrices $\mathbf{X}_{aug}$ and $\mathbf{\Psi}$, as both IQ imaging and nonlinear distortion will then be taken into account. Thus, let us define
\begin{align}
	\mathbf{\Psi}_{aug} = \begin{bmatrix} \mathbf{X}_{aug} & \mathbf{\tilde{\Psi}}\end{bmatrix} \text{,} \label{eq:psi_aug}
\end{align}
where $\mathbf{X}_{aug}$ is as discussed in Section~\ref{sec:wl_canc}, and $\mathbf{\tilde{\Psi}}$ is as discussed in Section~\ref{sec:nl_canc}, but with the block corresponding to $p = 1$ removed. This is done to avoid having two coefficients for the linear term, as both $\mathbf{X}_{aug}$ and $\mathbf{\Psi}$ obviously include the original linear signal. Again, for a more detailed explanation for the inner structure of these basis matrices, we refer the reader to \cite{Korpi133} and \cite{Anttila13}. The estimation of the parameters can again be done similar to \eqref{eq:h_est} and \eqref{eq:nl_canc}, but now the convolution matrix is replaced by $\mathbf{\Psi}_{aug}$. Using the established notation of Sections~\ref{sec:wl_canc} and~\ref{sec:nl_canc}, the resulting parameter vector consists of the estimates for $h_1(n)$, $h_2(n)$, and $f_{p,\mathit{eff}}(n)$ but with $p \neq 1$. Using these estimates, a cancellation signal can then be formed by utilizing the corresponding basis matrix.

It should be noted that this type of a joint cancellation scheme is not strictly speaking optimal in the sense that it does not include, for instance, the receiver image components of the nonlinearly distorted SI signal. Thus, it is not possible to perfectly reconstruct the observed SI signal using this model. Nevertheless, we will next show with waveform simulations that this scheme can significantly increase the achievable SINR, as it will still consider the most dominant distortion components. Developing a more comprehensive joint digital cancellation scheme is left for future work.

\section{Waveform Simulations}
\label{sec:simul}
\vspace{-0.5mm}

To evaluate the performance of the aforementioned enhanced SI cancellation algorithms, we will perform waveform simulations. A similar transceiver model, as shown in Fig.~\ref{fig:block_diagram}, is simulated and all the impairments discussed in this paper will be taken into account. Also the effect of TX-induced thermal noise is included in the simulations, unlike in any of the previous studies. Thus, the obtained results will provide a realistic estimate for the achievable performance with the different cancellation algorithms. The parameters presented in Tables~\ref{table:system_parameters} and~\ref{table:rx_parameters} are used also in the simulations, alongside with the additional parameters shown in Table~\ref{table:simul_param} which specify the utilized OFDM waveform. Note that the K-factor of the SI channel given in Table~\ref{table:simul_param} is based on actual measurements \cite{Duarte12}. A K-factor of this magnitude also indicates that 30 dB of RF cancellation is a realistic assumption.

\begin{table}[!t]
\renewcommand{\arraystretch}{1.3}
\caption{Additional parameters for the waveform simulator.}
\label{table:simul_param}
\centering
\begin{tabular}{|c||c|}
\hline
\textbf{Parameter} & \textbf{Value}\\
\hline
Constellation & 16-QAM\\
\hline
Number of subcarriers & 64\\
\hline
Number of data subcarriers & 48\\
\hline
Guard interval & 16 samples\\
\hline
Sample length & 15.625 ns\\
\hline
Symbol length & 4 $\mu$s\\
\hline
Signal bandwidth & 12.5 MHz\\
\hline
Oversampling factor & 4\\
\hline
K-factor of the SI channel & 35.8 dB\\
\hline
\end{tabular}
\vspace{-5.5mm}
\end{table}


The actual evaluations are done by determining the achieved SINRs from the simulations for four different digital SI cancellation algorithms: widely-linear cancellation presented in Section~\ref{sec:wl_canc}, nonlinear cancellation presented in Section~\ref{sec:nl_canc}, joint augmented nonlinear cancellation presented in Section~\ref{sec:joint_canc}, and traditional linear cancellation. In the simulations, each channel estimate is calculated during a calibration period when there is no actual signal of interest present, using 10000 samples, and the length of the individual channel response estimates ($M$) being set to 10. This means that, in widely-linear, nonlinear, and joint cancellation, both $\hat{h}_1(n)$ and $\hat{h}_2(n)$, alongside with all $f_{p,\mathit{eff}}(n)$, are of that length. Furthermore, the highest nonlinearity order, i.e., the parameter $P$, is set to 5, as a 5th-order model for the PA is used in the simulations. To ensure accurate results, the PA model is also set to have some memory.

\begin{figure}[!t]
\centering
\includegraphics[width=\columnwidth]{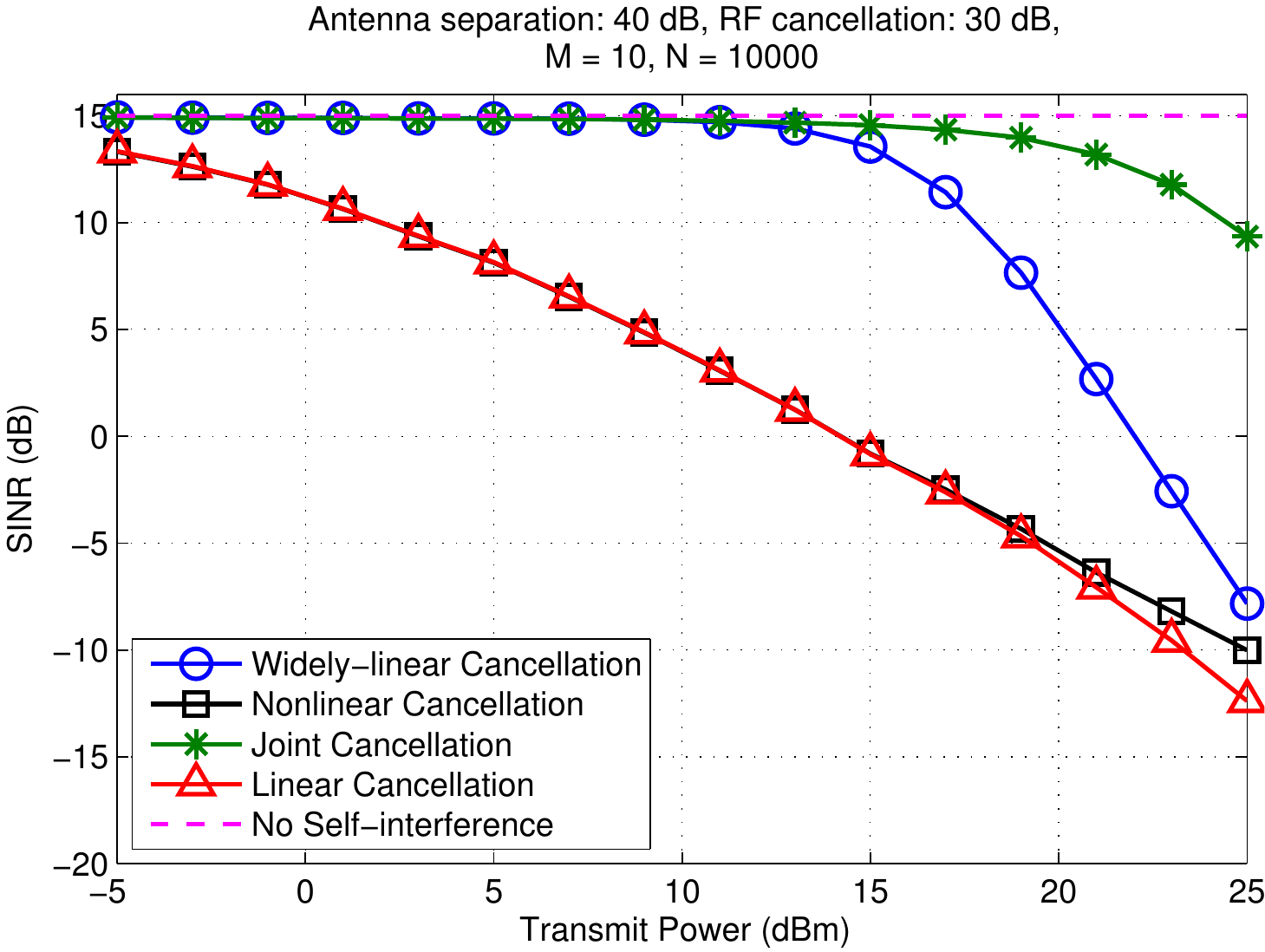}
\caption{The simulated SINRs for the different digital cancellation algorithms.}
\label{fig:sinrs}
\vspace{-5mm}
\end{figure}

The resulting SINR curves are shown in Fig.~\ref{fig:sinrs}, alongside with a reference curve correponding to a case without any self-interference. It can be observed that, with these RF component specifications, widely-linear digital cancellation clearly outperforms both nonlinear and linear cancellation. It actually achieves the SI-free SINR of 15 dB with transmit powers below 15 dBm, while neither linear nor nonlinear cancellation can achieve it even with the lowest considered transmit power. With transmit powers above 15 dBm, PA-induced nonlinear distortion starts to then limit the performance of widely-linear cancellation, as it is not able to take into account any nonlinear transformations of the signal. Note that this case is not directly comparable to that presented in \cite{Korpi133} as a more linear PA is assumed there.

On the other hand, Fig.~\ref{fig:sinrs} also indicates that the gain of performing nonlinear digital cancellation instead of the traditional linear cancellation is very minor in this case. It improves the SINR only marginally with the highest transmit powers. Thus, even though the PA-induced nonlinearities affect the SINR after widely-linear cancellation, it is evident that, with an IRR of 30 dB for both the TX and RX chains, conjugate SI is the dominant source of distortion.

Based on these observations, it is not surprising that the best performance is achieved with the proposed joint cancellation scheme. It clearly outperforms even widely-linear cancellation, as its performance is not limited by the PA-induced nonlinearities. Thus, even this type of a simple joint cancellation scheme can significantly improve the achievable SINR of a practical in-band full-duplex transceiver. The limiting factors for the SINR are now the image and nonlinear components not included in the model defined by \eqref{eq:psi_aug}, as well as TX-induced thermal noise, and nonlinear distortion produced by the receiver components. Our future work will concentrate on improving the joint cancellation scheme by including some of these aspects into the modeling, in particular the receiver nonlinearities.



\vspace{-2mm}
\section{Conclusion}
\label{sec:conc}
\vspace{-1mm}

In this paper, we have provided an overview on the different RF impairments occurring in in-band full-duplex transceivers. In addition, possible methods for compensating them were also discussed. Unlike in previous literature, also the effect of transmitter-induced thermal noise was taken into account in the analysis and simulations. Overall, it was shown that IQ imaging and transmitter power amplifier induced nonlinearities are some of the most considerable impairments in in-band full-duplex transceivers. We also evaluated and compared the performance of several different algorithms for compensating these impairments, and it was observed that they improved the SINR over traditional linear processing methods. The highest performance was achieved with a novel joint augmented nonlinear canceller, which is able to model both IQ imaging and nonlinear distortion. Our future work will concentrate on improving this proposed joint cancellation algorithm to further increase the efficiency of digital self-interference cancellation.

\vspace{-2mm}
\bibliographystyle{./IEEEtran}
\bibliography{./IEEEabrv,./IEEEref}

\end{document}